\documentclass[9pt,twocolumn,twoside]{pnas-new}
\usepackage{subcaption}
\templatetype{pnasresearcharticle}

\makeatletter
\let\PNAS@mk@linecount\relax
\let\PNAS@mk@linecountsecpage\relax

\def\PNAS@linecountLO{}
\def\PNAS@linecountRO{} 
\def\PNAS@linecountLE{}
\def\PNAS@linecountRE{}
\def\PNAS@linecountsecpageLO{}
\def\PNAS@linecountsecpageRO{}
\def\PNAS@linecountsecpageLE{}
\def\PNAS@linecountsecpageRE{}

\fancypagestyle{firststyle}{
   \fancyfoot[R]{\footerfont PNAS\hspace{7pt}|\hspace{7pt}\textbf{\today}\hspace{7pt}|\hspace{7pt}vol. XXX\hspace{7pt}|\hspace{7pt}no. XX\hspace{7pt}|\hspace{7pt}\textbf{\thepage\textendash\pageref{LastPage}}}
   \fancyfoot[L]{\footerfont\@ifundefined{@doi}{}{\@doi}}
   \fancyhead[LO]{}
   \fancyhead[RO]{}
   \fancyhead[LE]{}
   \fancyhead[RE]{}
}

\fancypagestyle{plain}{
   \fancyfoot[LE]{\footerfont\textbf{\thepage}\hspace{7pt}|\hspace{7pt}\@ifundefined{@doi}{}{\@doi}}
   \fancyfoot[RO]{\footerfont PNAS\hspace{7pt}|\hspace{7pt}\textbf{\today}\hspace{7pt}|\hspace{7pt}vol. XXX\hspace{7pt}|\hspace{7pt}no. XX\hspace{7pt}|\hspace{7pt}\textbf{\thepage}}
   \fancyfoot[RE,LO]{\footerfont\@ifundefined{@leadauthor}{}{\@leadauthor}\ifnum \value{authors} > 1\hspace{5pt}\textit{et al.}\fi}
   \fancyhead[LO]{}
   \fancyhead[RO]{}
   \fancyhead[LE]{}
   \fancyhead[RE]{}
}
\makeatother

\begin{document}

\title{ChromRec: Self-Assembly of Nucleosomes Driven by Directional Recognition}

\author[a]{Hesam Arabzadeh}
\author[a]{Dmitri Kireev}

\affil[a]{Department of Chemistry, University of Missouri, Columbia, Missouri 65211-7600, USA}

\leadauthor{Arabzadeh}

\significancestatement{Chromatin’s organization controls which genes turn on or off, but most experimental methods provide only static snapshots and most computer models either zoom in too far or zoom out too much. We introduce a new simulation framework that bridges this gap by encoding atomic‑level “molecular recognition” into protein‑scale beads. With it, we show that histone subunits spontaneously assemble into octamers and that DNA stays correctly wrapped around them, even in crowded mixtures. This capability allows, for the first time, dynamic genome‑scale simulations that still preserve key molecular details. Such multiscale modeling will help clarify how chromatin structure guides gene regulation and will inform strategies for epigenetic editing and disease therapy.}

\authordeclaration{The authors declare no competing interest.}
\correspondingauthor{\textsuperscript{2}To whom correspondence should be addressed. E-mail: dmitri.kireev@missouri.edu}


\begin{abstract}
Understanding chromatin dynamics across multiple spatiotemporal scales requires models that reconcile biological specificity with physics-based interactions and computational tractability. We present a modular, recognition-enabled ultra-coarse-grained (UCG) framework that captures both histone-DNA and histone-histone interactions using site-specific, off-center “Recognition” potentials. These \textit{recognition} sites, combined with generic attractive and repulsive terms, encode directional and stoichiometrically faithful assembly rules. Benchmark simulations demonstrate that this scheme robustly drives the self-assembly of geometrically correct histone octamers and enables stable nucleosome formation. The model also supports tunable resolution, allowing simplification of intra-octamer, nucleosomal, or fiber-level structures depending on the biological question to be addressed. This flexibility is especially useful for exploring chromatin reorganization driven by epigenetic regulation. While developed with chromatin in mind, our framework generalizes to other multivalent assemblies governed by molecular recognition.
\end{abstract}

\dates{This manuscript was compiled on \today}
\doi{\url{www.pnas.org/cgi/doi/10.1073/pnas.XXXXXXXXXX}}

\maketitle
\thispagestyle{firststyle}
\ifthenelse{\boolean{shortarticle}}{\ifthenelse{\boolean{singlecolumn}}{\abscontentformatted}{\abscontent}}{}

\firstpage[4]{3}


Chromatin's intricate architecture and dynamic role in gene regulation have been at the forefront of molecular biology research for decades. Proper chromatin organization is essential for maintaining genomic stability and regulating gene expression, with its deregulation implicated in various diseases, including cancer \cite{gardner2011operating,keenan2021discovering,black2012histone,loh2012actors,smith2013dna,bulger2011functional,hnisz2013super,calo2013modification}. For instance, mutations in chromatin remodeling complexes can disrupt normal cell identity and activate oncogenic phenotypes, contributing to cancer development \cite{schaefer2023roles,zhao2021language}. Understanding chromatin dynamics is crucial for developing medical applications, such as regenerative medicine and cancer therapies \cite{feng20213d,apostolou2013chromatin}. However, experimental methods like Hi-C \cite{stevens20173d}, ChIA-PET \cite{tang2015ctcf}, 3D-FISH \cite{markaki2012potential}, ChromEMT \cite{ou2017chromemt} primarily provide static snapshots of chromatin structure, lacking the temporal resolution to capture dynamic processes. While techniques like Hi-C offer averaged interactions over time, they do not reveal real-time dynamics \cite{stevens20173d}. Computational simulations have become indispensable for modeling chromatin dynamics, allowing researchers to explore temporal changes and predict functional outcomes. Although the experimental techniques may not be the best or only options in probing chromatin dynamics, they provide the most accurate data for calibrating simulations and validating their results \cite{lieberman2009comprehensive}.

Numerous computational models, particularly ultra-coarse-grained (UCG) simulations, have been developed to study chromatin organization and epigenetic phenomena at scales beyond the reach of experiments \cite{lequieu20191cpn,macpherson2018bottom,falk2019heterochromatin,bascom2018mesoscale,brackley2016simulated,brandani2021kinetic,zhang2016exploring}. Models based on the worm-like chain representation, for example, have successfully simulated chromatin fiber dynamics at the multi-mega-base scale, providing valuable insights into large-scale chromatin organization \cite{macpherson2018bottom,laghmach2020mesoscale,laghmach2021interplay,michieletto2018shaping,shi2018interphase,verdaasdonk2011centromeres,vasquez2016entropy,gursoy2016three,buckle2018polymer,michieletto2016polymer,kang2015confinement}. However, while these models excel in capturing macroscopic features, they often sacrifice resolution, leaving critical processes at smaller scales, such as transcription factor binding, histone modifications, and enhancer-promoter interactions, poorly understood. On the other hand, although models with higher resolution \cite{bascom2018mesoscale,brandani2021kinetic,zhang2016exploring,arya2006role,ozer2015chromatin} can capture microscopic events happening in the system, they cannot provide a larger picture of chromatin dynamics at the mega-base scale which is due to the computational cost. The absence of a model that bridges these scales highlights a gap in our ability to study multi-scale processes in gene regulation.

Addressing this gap is critical because gene expression relies on both large-scale chromatin organization and highly localized molecular interactions. For instance, distal enhancer-promoter interactions often span megabase regions, yet local variations in chromatin landscape or transcription factor (TF)-binding loci can have significant regulatory effects \cite{mendenhall2013locus,benveniste2014transcription}. Chromatin simulations at a sub-nucleosome resolution in megabase-scale systems would be important for better understanding the gene regulation mechanisms and inform the development of novel therapeutic modalities. In particular, recent advances in CRISPR/dCas9-based epigenetic editing gave rise to epigenetic transcription modifiers (ETM), new therapies that recruit epigenetic factors (e.g., writers or readers of histone marks) to specific genomic sites, thereby altering chromatin states and gene expression \cite{laghmach2020mesoscale,laghmach2021interplay,michieletto2018shaping,shi2018interphase}. Minor shifts in the DNA-binding locus by just a few base pairs can dramatically change the ETM efficacy, underscoring the necessity of a model with both high resolution and large-scale capabilities \cite{stanton2018chemically,zhou2024induced,hathaway2012dynamics,chiarella2020dose}.

High-resolution simulation of chromatin at genome scale must reconcile two opposing requirements: representing atomistically specific molecular recognition while maintaining tractable dynamics over biologically relevant times. Many chromatin processes rely on atomic-level readout -- e.g., recognition of lysine methylation states by reader domains and assembly of heterogeneous complexes such as PRC1/2 -- pushing toward atomistic detail. Yet, increasing resolution drives a polynomial rise in cost because finer beads increase particle count and require smaller timesteps (and hence a greater number of steps to reach desired simulated trajectories). We resolve these competing requirements by (i) introducing a family of off-center \textit{recognition} potentials that encode partner- and context-specific recognition while enforcing binding geometry, and (ii) using GPU-accelerated LAMMPS code to maximize throughput. The \textit{recognition} potential term activates only for designated partner types in close proximity, and the off-center positioning creates directional patches that prevent indiscriminate aggregation, thus preventing uncontrolled aggregation and allowing a single protein bead to engage multiple partners with defined orientation. This design preserves atomic-level recognition logic while using protein-scale beads and time steps on the order of $1~ns$, enabling access to biologically relevant trajectory lengths. Although the approach requires that specific pairwise interactions be specified a priori -- precluding discovery of unknown interaction partners -- the rapid expansion of curated protein-protein interaction data and increasingly accurate structure predictions (e.g., by AlphaFold3) render this constraint less and less limiting. By encoding these pairwise interactions in the force field, genome-scale simulations can reveal mechanisms of collective phenomena not accessible to direct experimental observation.

To demonstrate the model’s capabilities, we deliberately select challenges that demand higher-order, geometry-constrained recognition. We simulate (i) de novo self-assembly of the histone octamer from individual histone subunits initialized as randomized, non-native mixtures, and (ii) recognition of the histone octamer by a double-stranded DNA. These benchmarks require native stoichiometry, mutually consistent orientations, and directional protein-DNA contacts -- behaviors that conventional physics-based, ultra-coarse-grained models struggle to realize. Together they serve as stringent stress tests, demonstrating that \textit{recognition} potentials at protein-bead resolution can encode atomistic recognition logic. Establishing this capability lays the methodological foundation toward chromosome-scale simulations within the same framework.

\section*{Results}
\subsection*{Model}\

\begin{figure}[!t]
    \centering

    \begin{subfigure}[t]{0.2\textwidth}
        \includegraphics[width=\linewidth]{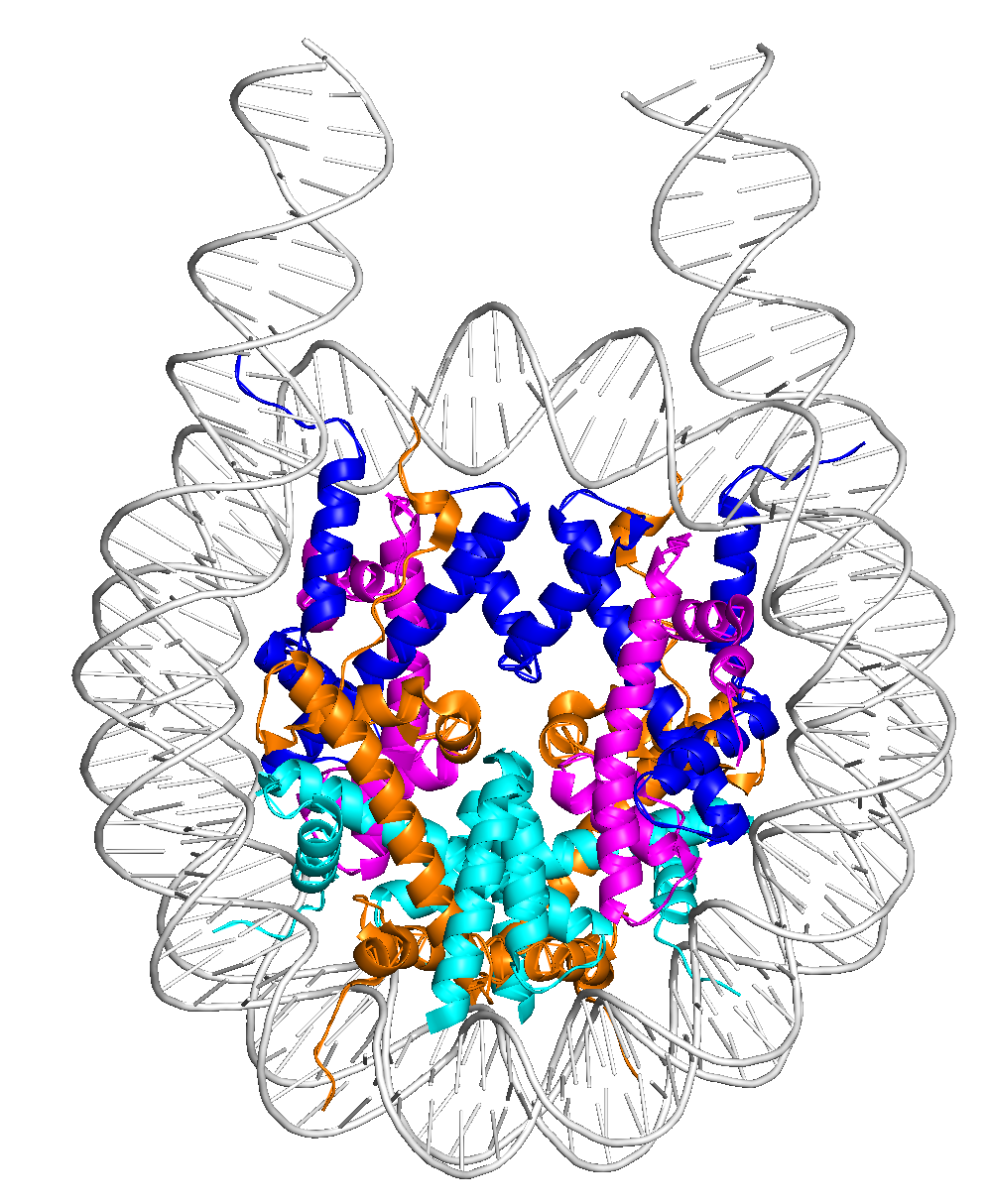}
        \caption{}
         \label{fig:nuc_front}
    \end{subfigure}
    \hspace{0.5em}
    \begin{subfigure}[t]{0.2\textwidth}
        \includegraphics[width=\linewidth]{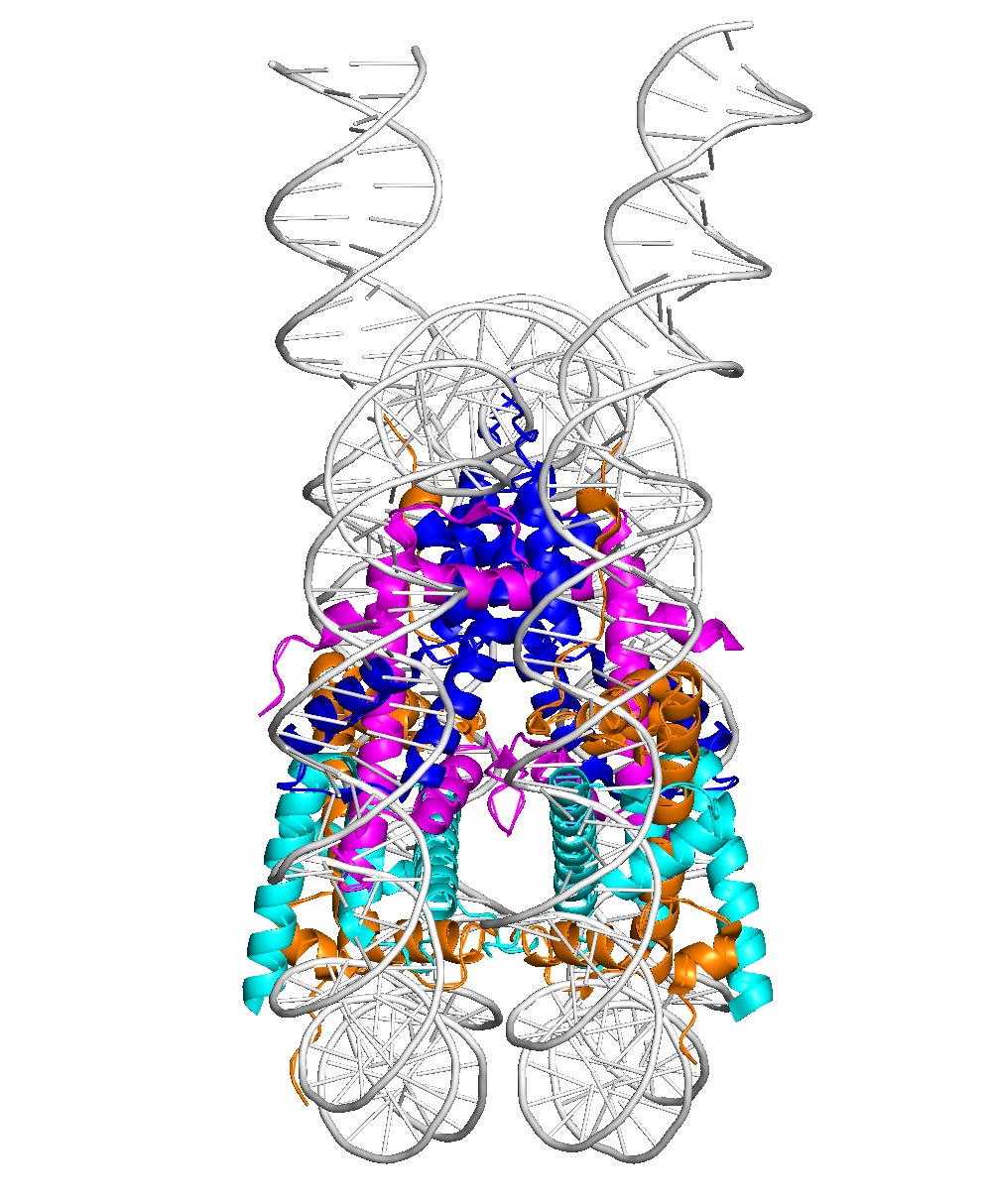}
        \caption{}
        \label{fig:nuc_side}
    \end{subfigure}

    \begin{subfigure}[t]{0.18\textwidth}
        \includegraphics[width=\linewidth]{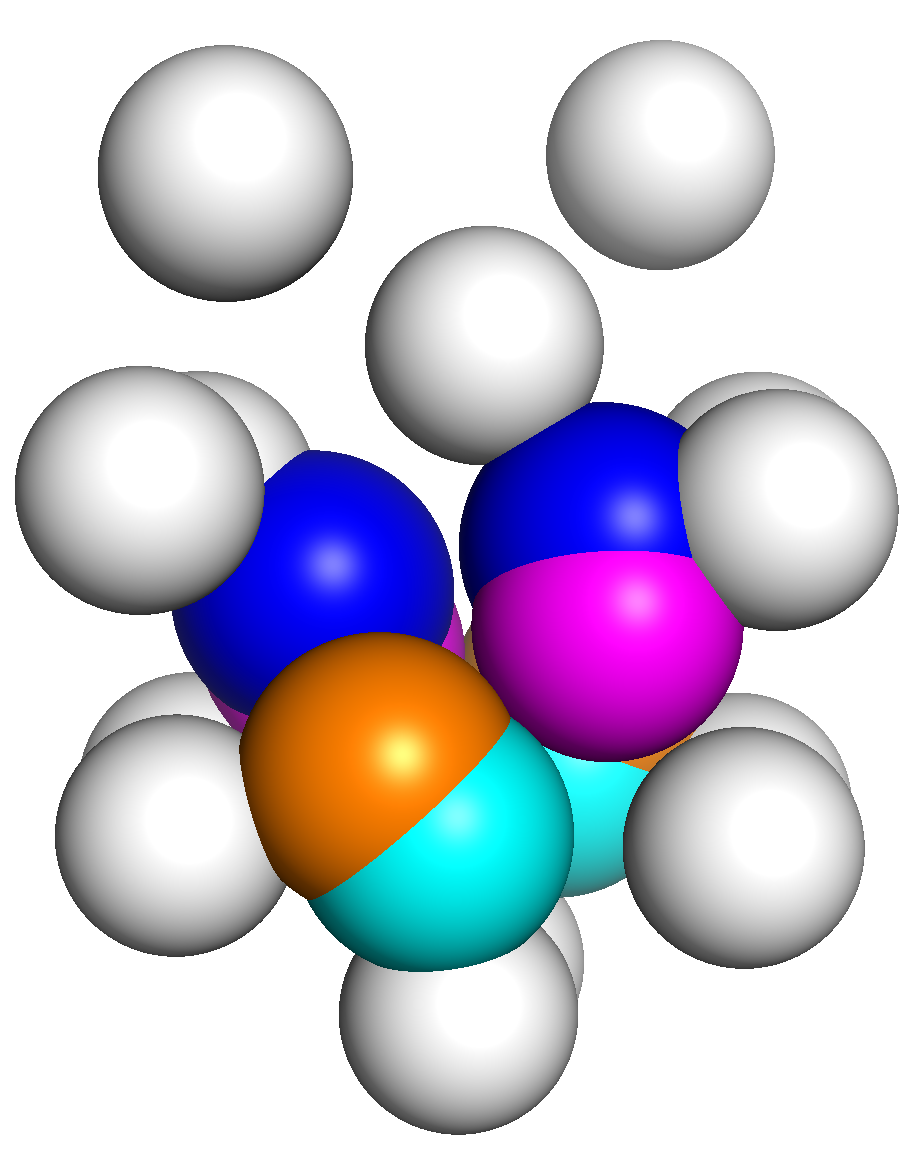}
        \caption{}
        \label{fig:cg_front}
    \end{subfigure}
    \hspace{0.82cm} 
    \begin{subfigure}[t]{0.14\textwidth}
        \includegraphics[width=\linewidth]{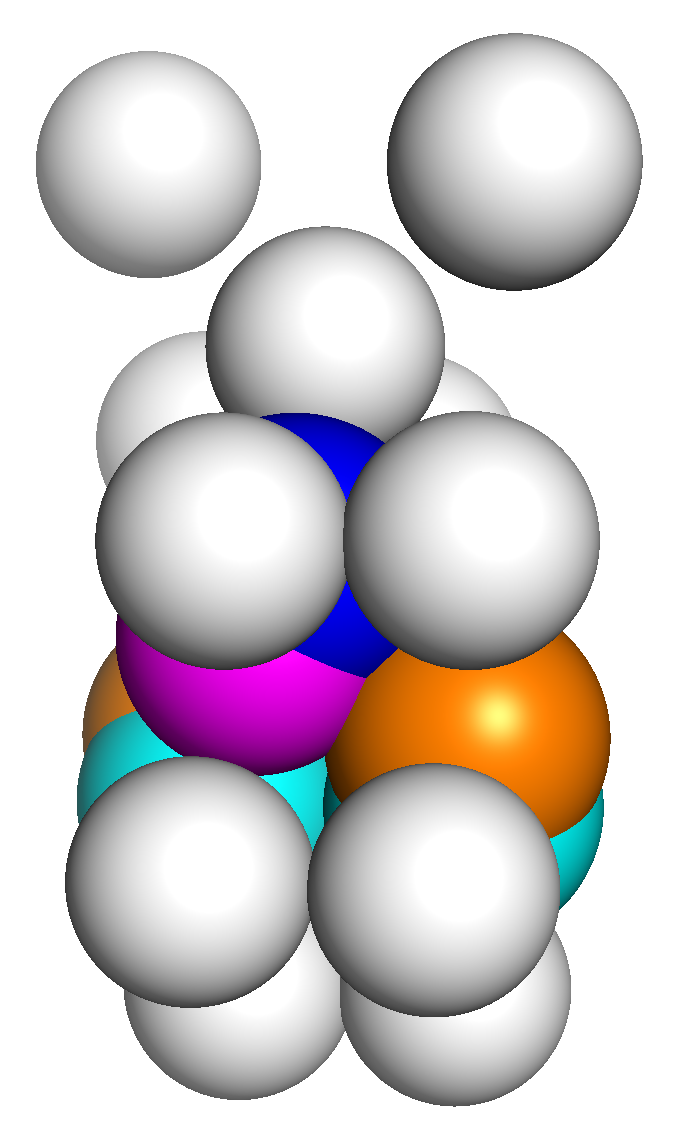}
        \caption{}
        \label{fig:cg_side}
    \end{subfigure}

    \begin{subfigure}[t]{0.28\textwidth}
        \includegraphics[width=\linewidth]{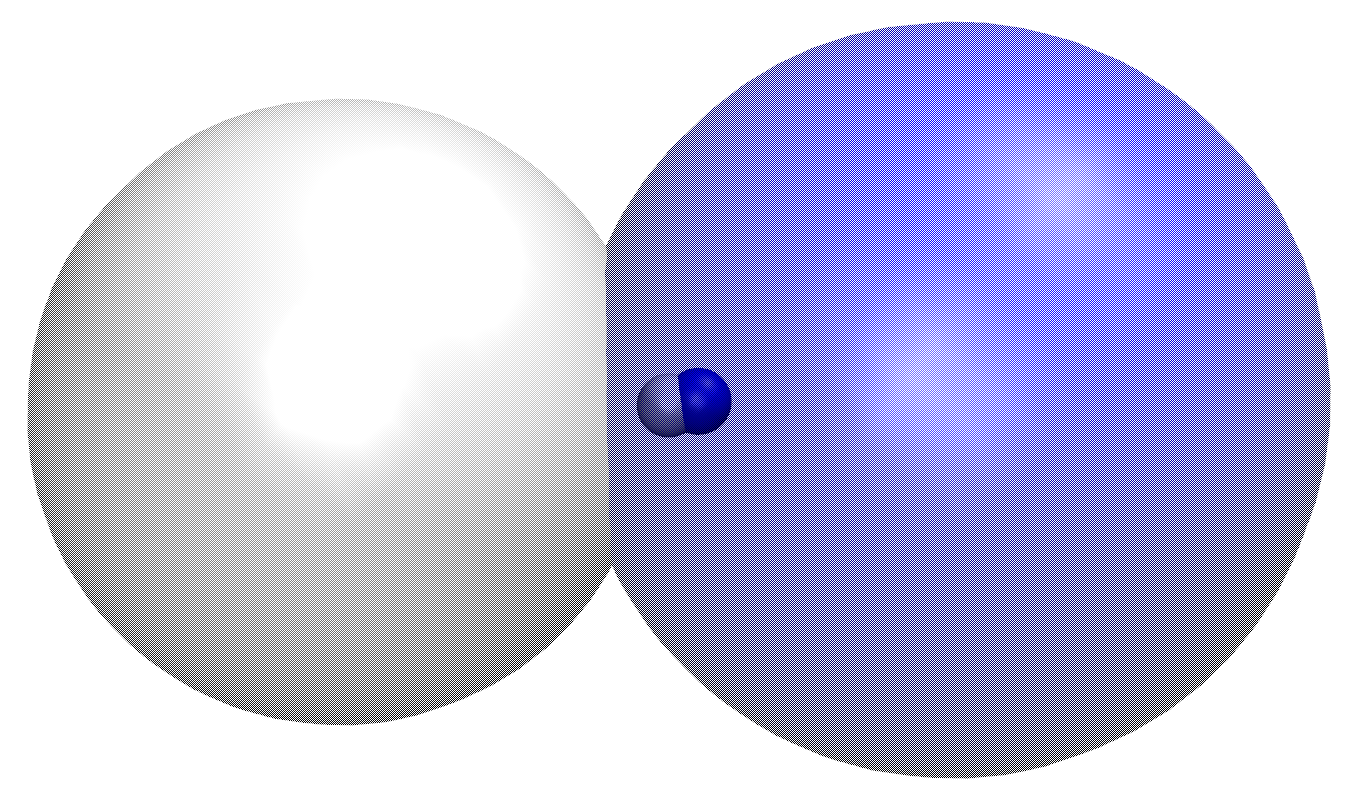}
        \caption{}
        \label{fig:site}
    \end{subfigure}

    \caption{\textbf{A)} An X-ray structure of human nucleosome (PDB ID=7xzy);
             \textbf{B)} Ultra-coarse-grained nucleosome representation as spheres with effective radii;
             \textbf{C)} \textit{Recognition} sites (\textit{small spheres}) for the histone H3 (blue) and DNA (white) beads.}
\end{figure}

Our model pursues protein-scale resolution while retaining atom-scale molecular recognition. Each histone subtype (H3, H4, H2A, H2B) is represented by a single protein bead; DNA is a semiflexible polymer with $\sim13$ bp per bead (slightly more than one helical turn), yielding eleven beads for the nucleosomal segment (see examples of X-ray and ultra-coarse-grained nucleosome models in Figure \ref{fig:nuc_front}-\ref{fig:cg_side}). “Bead size” should not be read as a hard-sphere radius but as an effective parameter setting the range of diffuse excluded-volume and other nonbonded terms. We employ two different octamer representations in the two case studies. For histone self-assembly, the four histone subtypes are simulated as separate beads endowed with off-center \textit{recognition} sites, and octamers emerge de novo under these interactions. For nucleosome stability, the octamer is treated as a preassembled hetero-oligomer: histone beads are held in native geometry by harmonic bonds (and angles), while DNA engages the same recognition scheme. This bonded octamer representation is also the default for chromosome-scale applications, ensuring octamer stability and reducing the number of evaluated nonbonded interactions.

\textbf{Force field.} Proteins are irregular and conformationally dynamic; spherically symmetric atomic Lennard-Jones (LJ) forms are therefore too sharp at this resolution. We instead use (i) a diffuse attractive term and (ii) a soft repulsive term to capture the collective effect of many atomic contacts without imposing hard-sphere behavior, together with (iii) an electrostatic term between charged particles and (iv) a \textit{recognition} potential that encodes partner- and orientation-specific binding. The attractive and repulsive interactions are isotropic at the bead level but broadened relative to atomic LJ to reflect protein shape heterogeneity; the bead “radius” should thus be interpreted as a mere parameter of the repulsive term, not a literal physical radius. Although the electrostatic term is part of the general force field, it is not used in the two case studies presented here, but can be included in chromosome-scale simulations. Full analytic expressions, combining rules and force field parameterization are given in Methods.

\textbf{DNA model.} DNA beads are connected by harmonic bonds and angles to form a semiflexible polymer. Parameters were obtained in two steps: a bottom-up fit to an atomistic MD trajectory of a complete nucleosome  \cite{armeev2021histone}, followed by a top-down refinement to reproduce the DNA persistence length. Using bond-direction vectors between consecutive beads, we fit the exponential decay of the orientational correlation along the contour to target a persistence length of $\sim45~nm$, within the accepted $40-50 nm$ range (see Methods for details).

\textbf{Recognition potential.} Specific protein-protein and protein-DNA recognition is implemented by off-center \textit{recognition} sites tethered to their parent protein bead by harmonic springs. These sites carry no repulsive or electrostatic terms; they interact only via a partner-gated attractive potential that activates within a prescribed distance-and-orientation window. Several sites per protein bead allow controlled multivalency (e.g., distinct H3-H4, H3-H2A, H3-H2B contacts) while enforcing geometry and preventing indiscriminate aggregation (Figure \ref{fig:site}). Site locations and intra-bead angular restraints are initialized with the nucleosome crystal structure (PDB 7XZY) and maintained by harmonic angle terms. The geometry of \textit{recognition} sites is illustrated in Figure \ref{fig:site}. \textit{Recognition} potentials were calibrated in a top-down fashion from prior biophysical measurements \cite{kabsch1976solution,hathaway2012dynamics} as detailed in Supplementary Information.

\textbf{Dynamics.} Trajectories are propagated under overdamped Langevin dynamics at temperature T with friction $\gamma$ and time step $dt$ (of $1~ns$). The stochastic thermostat provides solvent-like damping and thermal noise while preserving the recognition-driven assembly pathways. The full overdamped Langevin equation and parameter values are specified in Methods.

\subsection*{Self-assembly of the histone octamer}\

The histone octamer, a protein core comprising two copies each of H2A, H2B, H3, and H4, is the fundamental unit around which DNA wraps to form nucleosomes. A key test of our ultra-coarse-grained force field is whether these eight histone subunits can self-assemble from random initial conditions into native-like octamers solely under the influence of partner- and geometry-specific \textit{recognition} potentials. To evaluate this, we conducted a series of simulations in increasing complexity: (i) de novo formation of a single octamer under ideal stoichiometry, (ii) negative-control runs with incorrect or incomplete histone compositions, and (iii) competitive assembly from a crowded mixture of 256 histone subunits.

\textbf{Single-octamer assembly.} The primary goal here is to verify whether our recognition-based force field can robustly drive de novo self-assembly of histone octamers from randomized initial conditions. Specifically, we aim to test whether eight histone subunits, two copies each of H3, H4, H2A, and H2B, can spontaneously organize into the correct structure under our partner- and orientation-specific interaction. The resulting self-assembled octamer should reproduce the structure, geometry, and mutual positioning of histone subtypes observed in the X-ray nucleosome structure. To assess structural correctness, once the octamer is assembled, we compared it to the native crystal geometry using the Kabsch RMSD algorithm throughout the entire trajectory, which identifies the optimal alignment between two point sets under all possible subtype label permutations. In particular, the algorithm searches for the best-matched histone types in the simulation relative to the native geometry (crystal structure) by computing the optimal rotation matrix that minimizes the RMSD between them. This permutation step is essential due to the symmetry of the system: any of the two H3's in simulation could correspond to either H3 in the crystal structure, and the same holds for the other subtypes. We performed 50 independent simulations, each initialized with randomly positioned and oriented histone beads in a cubic simulation box. The system included two copies of each histone subtype, consistent with the stoichiometry of the native octamer. No positional restraints or bonded interactions were applied, so the octamer was expected to form purely via nonbonded \textit{recognition} terms. A successful assembly was defined by both (i) formation of a compact 8-bead complex, and (ii) correct relative arrangement of subunits. The mean Kabsch RMSD across 50 simulations was $0.98~nm$, with an uncertainty $0.59~nm$ along a random baseline coefficient of $1.8~nm$. The random baseline was obtained by randomly reassigning histone subtype identifications 100 times for each frame across all 50 simulations and computing the Kabsch RMSD to the reference structure. The resulting average ($1.8~nm$) represents the expected RMSD for random configurations, providing a statistical reference showing that our assembled octamers ($0.98~nm$) are substantially more native‑like than chance. A complete octamer formed in all trajectories (see Movie S1 in Supporting Information for a representative run), with a mean assembly time of 435 $\mu s$. Visual inspection of a structure with the mean RMSD aligned with the native geometry  (see Supporting Figure S1) confirms that the correct spatial arrangement of all eight histones was achieved. To further analyze internal packing, we computed pairwise distances between all histone subunits and compared these to reference values. The histone-histone distance distributions (Supporting Figure S2) show mean values agreeing well with the crystal octamer (Table S1), indicating reproducible mutual positioning across runs. Although the RMSD values appear slightly large compared to atomic-scale simulations, they are reasonable at this ultra-coarse resolution. The minimum histone-histone distance available in the native octamer is approximately $\sim1.0~nm$, and thermal fluctuations can produce displacements of $\sim0.11~nm$ per bead per time step. Given eight independent beads in motion, the observed RMSD value is expected and reflects realistic flexibility rather than structural failure, as confirmed by the aligned simulated and native crystal structures shown in Supporting Figure S1. These results confirm that the \textit{recognition} potential is sufficient to drive spontaneous formation of topologically and geometrically correct octamers from randomized conditions.

\textbf{Negative-control compositions.} To test the specificity and selectivity of our \textit{recognition} potential, we performed negative-control simulations with incorrect histone stoichiometries. These systems were designed to mimic scenarios where octamer assembly is structurally or chemically unfavorable, thereby favoring nonspecific aggregation and spontaneous misfolding. The goal is to verify that the force field does not artificially drive the formation of compact octamer-like structures with invalid compositions. We designed two sets of simulations, each run across 10 independent replicas. In the first set, only H3-H4 dimers were present, omitting the H2A and H2B subunits entirely. In the second set, we introduced two copies of each homo-type (H3-H3, H4-H4, H2A-H2A, and H2B-H2B). These control groups systematically lack the full complement of interaction partners defined in the native octamer, and thus test whether nonspecific attraction or geometric constraints alone could suffice to prevent non-specific aggregation. In both control scenarios, no trajectory yielded a complete octamer (see representative Movie S2 in Supporting Information). Instead, the systems stalled at sub-octamer intermediates, such as dimers, and could not evolve into higher-order structures, in contrast to the robust octamer formation observed under correct stoichiometry. These negative results confirm that the recognition scheme does not promote spurious binding or nonfunctional aggregation. Octamer formation depends critically on both complementary interaction types and the correct stoichiometric balance. The \textit{recognition} potential is thus highly specific and selective, favoring correct assemblies over nonfunctional alternatives.

\begin{figure}[!t]
\centering
\includegraphics[width=.8\linewidth]{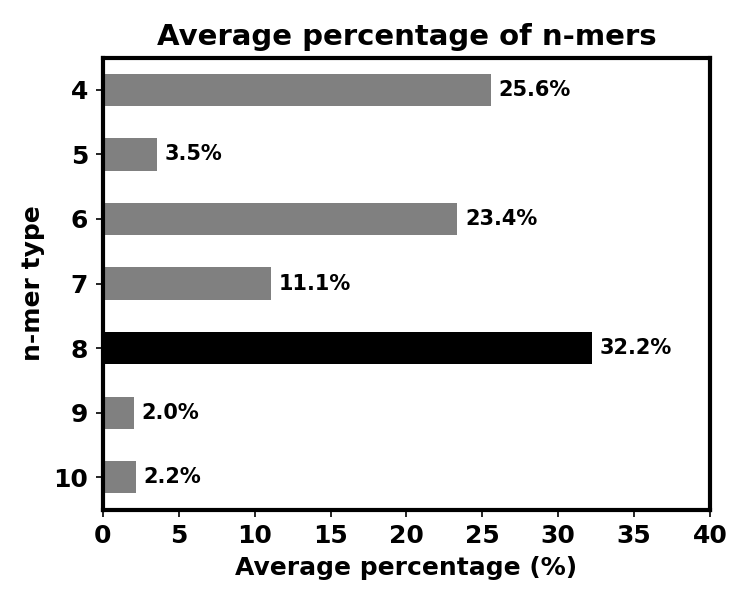}
\caption{Average percentages of N-mer complexes formed after a minute of simulation in five runs (gray bars) along with respective percentages of octamers involved (black bars).}
\label{fig:nmer}
\end{figure}

\textbf{Crowded-mixture assembly.} To assess the selectivity of our \textit{recognition} force field term under a biologically relevant regime, we challenged the system to assemble histone octamers within a crowded, heterogeneous environment. The goal of this test is to evaluate whether correct octamers can still form spontaneously when multiple competing pathways and intermediate species are available, as one would expect in cellular contexts. To quantify assembly progression, we tracked the abundance of each N-mer species (i.e., clusters containing N histone beads) as a function of time. We initialized a mixture of 256 histone beads, comprising 64 subunits of each subtype (H3, H4, H2A, H2B), randomly distributed throughout a cubic simulation box. This configuration corresponds to 32 possible octamers. Five independent replicas were run for $60~s$ of simulated time. No restraints or biasing forces were applied; octamer formation had to compete with the emergence of nonfunctional aggregates or incomplete intermediates. Monomers and small oligomers rapidly disappeared within the first few simulated seconds, replaced by a dominant population of tetrameric intermediates. By $\sim60~s$, fully formed octamers became the predominant species in four of five runs (see Figure \ref{fig:nmer}). A small number of higher-order aggregates (nonamers and decamers) were observed in some replicas, specifically, two nonamers in two runs, and decamers in three runs, indicating rare overgrowth events. Nevertheless, octamers consistently emerged as the thermodynamically favored endpoint. These results confirm that the \textit{recognition} potentials are sufficient to drive correct assembly even in the presence of numerous competing interactions and potential kinetic traps. The predominance of correctly sized octamers demonstrates not only the specificity of the \textit{recognition} potential but also its selectivity under high-concentration, crowded conditions. Together, these findings reinforce the conclusion that our force field promotes spontaneous, topologically correct assembly of higher-order protein complexes under complex biological scenarios.

\subsection*{Nuscleosome stability}\

This study was designed to evaluate the stability and structural fidelity of the nucleosome -- the fundamental unit of chromatin consisting of double-stranded DNA wrapped around a histone octamer. We aimed to determine whether \textit{recognition} potential alone is sufficient to maintain DNA in the correct wrapped conformation around the octamer and to assess its performance across multiple stochastic replicas. This setup represents a stress test for our force field, given the high stiffness of double-stranded DNA, whose persistence length greatly exceeds a single turn around the histone octamer. To isolate the contribution of DNA-histone interactions, the histone octamer was stabilized by applying harmonic bonds between subunits to preserve its internal geometry. DNA-histone interactions, however, were governed purely by the directional \textit{recognition} potential. Fifty independent simulations were initialized with different random seeds. No restraints were applied to the DNA beyond the \textit{recognition} term. Structural similarity to the native nucleosome geometry was evaluated using RMSD. In all fifty replicas, the nucleosome maintained its wrapped conformation over the course of the simulation, with a mean RMSD of $1.6~nm$ relative to the reference crystal structure. These results demonstrate that the \textit{recognition} potential can robustly maintain the nucleosomal architecture. To further test the active recognition-driven recruitment of DNA, we performed an additional set of ten simulations in which a bonded histone octamer and an initially unbound DNA strand were placed in the same simulation box at varying separations. In one of the simulations, a full assembly of the histone-DNA complex emerged spontaneously, a result that is particularly striking considering the system’s minimal constraints (see Movie S3). While the low success rate highlights kinetic challenges associated with spontaneous DNA wrapping, the one successful case illustrates that our potential can guide wrapping under favorable conditions. Finally, we tested the \textit{recognition} potential in a multi-nucleosome context by simulating a system composed of ten preassembled nucleosomes (with bonded octamers and DNA governed by recognition). The system remained structurally stable over the full trajectory, and a representative movie is available in the Supporting Information (Movie S4), demonstrating both the scalability and mechanical stability of our model. Together, these results confirm that the force field featuring \textit{recognition} potential is sufficient to preserve nucleosome structure under thermal fluctuations and across diverse initial conditions. 

\section*{Discussion}
We introduced a UCG chromatin framework that captures protein-protein and protein-DNA specificity while remaining scalable to genome contexts. Its key feature is an off-center \textit{recognition} potential: partner-gated interactions applied to tethered \textit{recognition} sites that enforce geometry and valence which otherwise could only be achieved at atomistic resolution.

Interestingly, anisotropic interactions are well established in soft-matter and colloid science, where “patchy” particles with directional bonding have been used to program self-assembly \cite{bucciarelli2016dramatic,zhu2020self,kraft2012surface}. Those models typically treat particles as hard bodies carrying surface patches. By contrast, proteins at a single-bead resolution are irregularly shaped, flexible objects that are better viewed as diffuse, entities whose average interactions are soft and orientation-dependent. We therefore implement anisotropy through the off-center \textit{recognition} sites tethered to protein beads, leaving excluded volume and electrostatics to the diffuse, isotropic terms. Surprisingly, to our knowledge, such recognition mechanisms have not yet been adopted for protein-scale recognition in ultra-coarse models of biological systems.

Our framework is open to tunable trade-off between resolution and throughput. For instance, when histone turnover or complex (dis)assembly \cite{tramantano2016constitutive} is of interest, each histone bead can carry \textit{recognition} sites, enabling association and dissociation of octamer interfaces under native stoichiometry. When the process of interest is downstream -- such as nucleosome sliding, eviction, or remodeler-mediated repositioning \cite{niina2017sequence} -- the histone octamer can be represented as a stable hetero-oligomer held by harmonic bonds, thus reducing degrees of freedom while preserving the non-bonded DNA-octamer recognition scheme. Furthermore, for studies focused primarily on fiber-level reorganization driven by changing Post-Translational Modification (PTM) landscapes \cite{bowman2014post}, even the DNA-octamer contacts can be represented by bonded terms, further increasing simulated time accessible per trajectory without altering the higher-level readout.

Our octamer self-assembly benchmarks demonstrate that the \textit{recognition} potential suffices to produce geometrically correct octamers from randomized mixtures under ideal stoichiometry and to reject incorrect compositions. The goal here was not to reproduce a unique assembly pathway; indeed, multiple routes were observed. However, if kinetic fidelity becomes important, the same \textit{recognition} term can be parameterized to favor experimentally supported routes so that the dominant pathways and intermediate lifetimes align with biochemical evidence \cite{brandani2021kinetic,hazan2015nucleosome}. In crowded mixtures, the emergence and dominance of octamers over time further support the robustness of the recognition scheme under competitive conditions.

Several limitations of our approach are worth noting. Because recognition is partner-specified, unknown interactions will not be discovered de novo; the approach is most powerful when informed by structural and interactome data which become more and more accessible. Hydrodynamics and explicit solvent are not included; their effects are subsumed into Langevin friction and noise. Finally, parameter choices (ranges, well depths, cutoffs) trade acuracy against speed; we report defaults that work across our benchmarks, but systematic calibration to target datasets will improve kinetic realism.

In conclusion, these benchmarks demonstrate how recognition-enabled, protein-scale dynamics can capture complex local chromatin behavior while remaining compatible with chromosome-scale simulations. Although developed with chromatin in mind, the same model should transfer to other recognition-governed assemblies-for example, clathrin-coated vesicle formation \cite{nawara2022imaging}, assembly and regulation of the transcription pre-initiation complex \cite{zhang2016rapid}, or multivalent scaffold-client interactions in nuclei \cite{banani2017biomolecular}. As larger genome-scale applications are reported, we anticipate that this framework will help connect molecular recognition logic to emergent, system-level chromatin behavior.







\matmethods{
\textbf{Coarse-grained particles.} Each particle in our system is treated as a dimensionless point that interacts through spherically symmetric potentials. To define a length scale for these interactions, we assign an effective radius to each particle. This effective radius enters into both attractive and repulsive (pairwise) potentials as the distance between interacting particles and is essential for computing radial separation. In particular, the repulsive potential uses this radius to control excluded volume and prevent unphysical overlap, especially relevant for histone pairs like H3-H4 and H2A-H2B. To determine the effective radii, we obtained the (PDB ID=7xzy) structure of a nucleosome from protein data bank, then considered the atomic radii plus solvent-accessible surface area (SASA). Then using a voxel-based algorithm with voxel length of 0.1 \AA~obtained their volume and finally calculated the radius from $4/3~\pi r^3$. Visual representation of a histone H3 with SASA plus its voxel-based depiction are provided in Figure S1. The final spheres overlaid on the atomic structure of a nucleosome are depicted in Figure 1. The derived radii (in $nm$) are H3 = 1.74, H4 = 1.62, H2A = 1.69, H2B = 1.67, and DNA = 1.44.

\textbf{Brownian dynamics, time step, and friction coefficient}. To reduce the degrees of freedom, including those associated with velocities and the solvent, and to account for the stochastic forces due to thermal fluctuations, the Brownian equations of motion (EOM) were employed:
\begin{equation}
\frac{dr}{dt} = -\frac{1}{\gamma}\nabla U(r) +  \sqrt{\frac{2 k_B T}{\gamma}} R(t), \label{eq:brownian}
\end{equation}

where $r$ is the position vector, $r = (x,y,z) \in \mathbb{R}^3$, $t$ is time, $\nabla U(r)$ is the deterministic force (potential gradient), $\gamma$ is the friction coefficient, $R(t)$ is the Gaussian white noise term, with $\langle R(t) \rangle = 0$ and $\langle R(t) R(t') \rangle  = \delta(t-t')$, $k_B$ is the Boltzmann constant, and $T$ is temperature. To be able to numerically integrate this stochastic differential equation, we use Euler--Maruyama algorithm which is a standard discretization technique. With timestep $\Delta t$, it gives: 
\begin{equation}
r(t+\Delta t) = r(t) - \frac{\Delta t}{\gamma}\nabla U(r(t)) + \sqrt{\frac{2 k_B T\Delta t}{\gamma}} \mathcal{N}(0, 1), \label{eq:EM}
\end{equation}
Within Brownian EOM two factors need to be addressed, the time step and the friction coefficient. First, the diffusion coefficient was calculated based on the viscosity of cell lysate, $\eta =~\sim0.2~\mathrm{Pa\cdot s}$ \cite{kong2009using}, where $Pa$ is the pressure unit, Pascal, $s$ is seconds, employing the Einstein-Stokes relation:
\begin{equation}
D_A  =  \frac{k_BT}{6\pi \eta r_A} = 6.3 \times 10^{-12}~\mathrm{m^2/s},
\end{equation}
where $D_A$ is the diffusion coefficient, and $r_A$ is the radius of particles, calculated in the previous step. Next, the root means square deviation (RMSD) of particles undergoing random diffusion were calculated for a few time-steps at nano scale and to have a stable simulation the time-step of 1 $ns$ was chosen:
\begin{equation}
S_A= \sqrt{2D_A \Delta t} = 1.1 \times 10^{-10}  m,
\end{equation} 
where $S_A$ is the RMSD of particles undergoing random diffusion, and $\Delta t (= 1 ns)$ is the time step. Next, the friction coefficient in Brownian equations of motion was varied to reproduce the required RMSD. To calculate the RMSD of the particles from the simulation trajectory, the mean displacement of the particles was calculated with respect to the previous step:
\begin{equation}
RMSD= \sqrt{\frac{1}{N} \sum_{i=1}^N \langle | r_i(t+\Delta t) - r_{i}(t) |^2 \rangle} ~,
\end{equation}
where $N$ is the number of particles, $r_i(t+\Delta t)$ is the position of particle $i$ at current and $r_{i}(t)$ is its position at previous steps.

\textbf{DNA force field and parameterization.} In our DNA model, a bond term (eq. \ref{eq:bond}) and an angle term (eq. \ref{eq:angle}) were employed both using harmonic springs:
\begin{align}
E_{bonded}(r_{ij}) = \frac{1}{2} K_{ij} (r_{ij} - r_{ij,~eq})^2 \label{eq:bond}\\
E_{angle}(\theta_{ijk}) = \frac{1}{2} K_{ijk} (\theta_{ijk} - \theta_{ijk,~eq})^2 \label{eq:angle}
\end{align}
where $r_{ij}$ and $r_{ij,~eq}$ are the distance and equilibrium distance between particles $i$ and $j$, respectively, $k_{ij}$ is the force constant associated with the bond between particles $i$ and $j$, $\theta_{ijk}$ and $\theta_{ijk,~eq}$ are the angle, and equilibrium angle between two segments, connecting vectors $i-j$ and $j-k$, respectively, the force constant $k_{ijk}$ corresponds to the angular interaction between particles $i$, $j$ and $k$. The equilibrium value of bonds and angles are acquired from their corresponding values in the crystal structure (PDB ID=7xzy). The bond and angle force constants were parameterized using data from the MD trajectory of ref \cite{shaytan2016coupling} employing the equipartition theorem for a harmonic potential. The force constants were calculated as, $\frac{k_B T}{\sigma(r)}$, $\frac{k_B T}{\sigma(\theta)}$, where $\sigma(r)$ and $\sigma(\theta)$ are associated with the variance of the bond lengths and angles, respectively, averaged over the entire MD trajectory. Next, the force constants were refined to reproduce the persistence length of a DNA $(\sim45~\mathrm{nm} / \sim450~\text{\AA})$. To do so, the bond direction vectors were calculated between consecutive particles, as the persistence length is associated with the angle between these bond vectors. Then, the correlation of these bond vectors was computed as a function of contour length, which is the cumulative distance along the polymer backbone. The correlation function (eq. \ref{eq:pers_leng}) was fit to an exponential decay \cite{kroon1997estimation}, and the decay length (Figure \ref{fig:pers_leng} 3) provided an estimate of the persistence length:
\begin{equation}
C(s) = \langle r_i \cdot r_{i+s} \rangle = \exp\!\left(-\frac{s}{l_p}\right), \label{eq:pers_leng}
\end{equation}
where $r_i$ and $r_{i+s}$ are bond vectors between consecutive particles at positions $i$ and $i+s$ along the polymer backbone, $s$ is the contour length of the polymer (DNA), and $l_p$ is the persistence length.

\begin{figure}[!t]
\centering
\includegraphics[width=.83\linewidth]{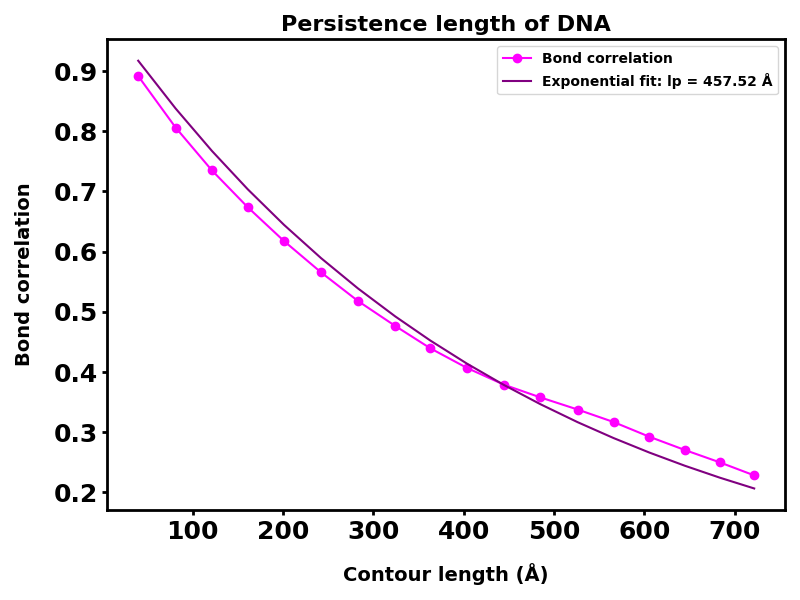}
\caption{Bond vector correlation vs contour length of DNA.}
\label{fig:pers_leng}
\end{figure}

\textbf{Histones’ force field and parameterization.} For \textit{recognition} interaction, the histones interact through site particles. In other words, it is the sites that keep the parent particles in an interaction. The parent histone particles are connected to their sites, that are point particles, through a harmonic spring. A site is positioned along a vector pointing from parent A to parent B at a fixed offset (center of them), as depicted in Figure \ref{fig:site}. To ensure rotational flexibility of the site particles, we tracked cumulative displacement vectors over time as a diagnostic for directional randomization. A converging vector sum was used as an indicator of rotational dynamics (see Figure S2), consistent with expectations that site motions are not fixed, and that rotational effects cause directional changes, leading to the cancellation of displacement vectors over time. For each parent histone particle three sites were considered for the interaction with the other histone sites. Though each histone could have four sites to be able to interact with all histone types, having three sites balances the computational efficiency and the stability of the histone octamer. For the parameterization of histone-histone interactions, an experimental value of micromolar-scale binding constant \cite{mardian1978yeast}, $k_d$, was employed. Considering $k_d = \frac{k_{off}}{k_{on}}$, having a micromolar-scale $k_d$ means a mean survival (or residence) time, $\tau$, of minute-scale, since $\tau= \frac{1}{k_{off}}$. To achieve this value, we carried out 20-minute-long simulations in triplicate for H3-H4, H2A-H2B, H4-H2A, H4-H2B, H3-H2A and H3-H2B dimers to be able to observe a mean residence time of minute-scale. Accordingly, the \textit{recognition}, attraction and repulsion terms have been optimized to obtain the mentioned residence time. Although reference \cite{mardian1978yeast} measured the binding constant for H2A-H2B, we parameterized all histone-histone interactions based on the same micromolar scale binding constant. Table S1 shows the mean survival time and mean binding constant of our simulations for each interaction. Then, in a step-by-step fashion, we added more particles to the dimers for finetuning the parameters in order to form and keep the octamer stable. In this step, we also took the distance between histones into consideration for forming a topologically correct structure of the nucleosomes. The distribution of radial distances of all histone-histone interactions are shown in Figure S2. The radial distances of the corresponding distances derived from the crystal structure are shown in Table S2. 

As mentioned earlier, the long-range potential terms of our UCG force field include, an attractive, a repulsive, and the \textit{recognition} potentials. The attractive term takes the form of a Gaussian function, providing a smooth and differentiable interaction landscape critical for accurate force and gradient calculations:
\begin{equation}
V_{attractive} = -A \exp\!\left(-\frac{(r_{dist} - r_{radius})^2}{2\sigma^2}\right) \label{eq:attract}
\end{equation}

\begin{figure}[!t]
\centering
\includegraphics[width=.8\linewidth]{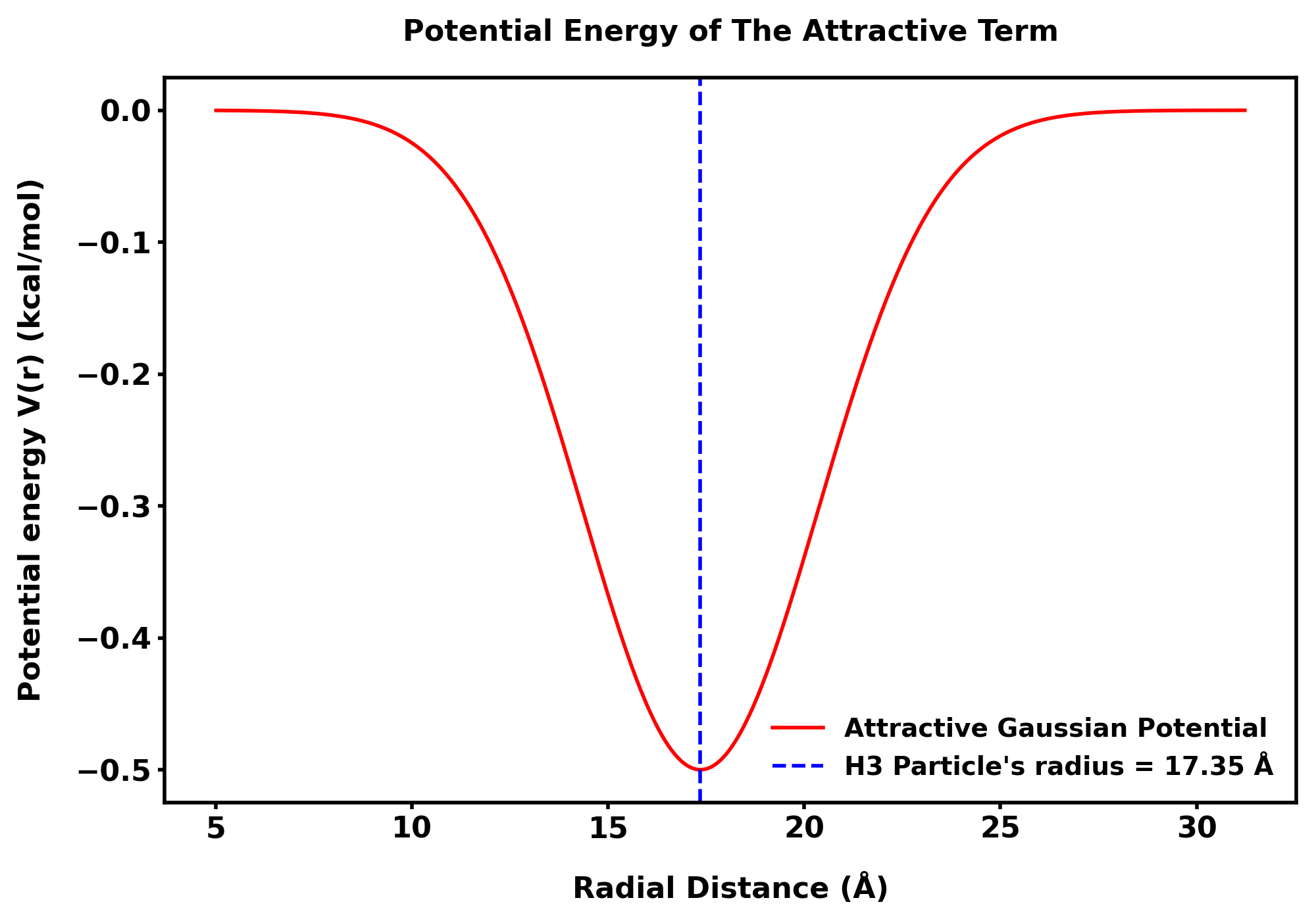}
\caption{Potential energy surface of the attractive term.}
\label{fig:attract}
\end{figure}

where $A$ represents the potential’s amplitude, $r_{dist}$ the distance between histone particles, $r_{radius}$ is the effective interaction radius assigned to each bead, and $\sigma$ the standard deviation, serving as a parameter for potential cutoff adjustment. The Lorentz--Berthelot combining rules were used in all potentials of our force field. The potential energy surface of the attractive term is shown in Figure \ref{fig:attract}. The repulsive term is defined by an exponential function,
\begin{equation}
V_{repulsive} = A \exp\!\left(-b(r_{dist} - r_{radius})\right) \label{eq:repuls}
 \end{equation}
 
 \begin{figure}[!t]
\centering
\includegraphics[width=.8\linewidth]{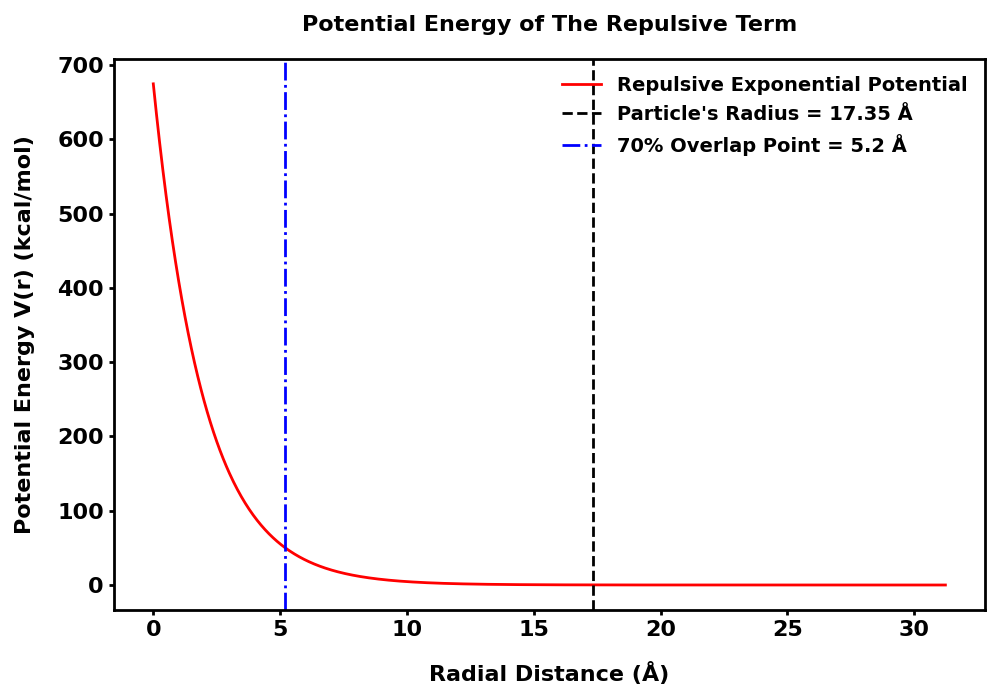}
\caption{Potential energy surface of the repulsive term.}
\label{fig:repuls}
\end{figure}

where $A$, $r$ and $d$ are the same as the attractive term with a difference in $r$ that accommodates the physical overlap observed in crystal structures for specific histone pairs, thereby enhancing the model’s structural accuracy, b the decay constant, accounting for the softness of the repulsion. The potential energy surface of the repuslive term is shown in Figure \ref{fig:repuls}. The \textit{recognition} potential again employs a Gaussian form with the difference that there is no radius considered here:
\begin{equation}
V_{Recognition} = -A \exp\!\left(-\frac{r_{dist}^2}{2\sigma^2}\right) \label{eq:recog}
\end{equation}
where all variables are similar to the previous potentials. The potential energy surface of the attractive term is shown in Figure \ref{fig:recog}.

 \begin{figure}[!t]
\centering
\includegraphics[width=.8\linewidth]{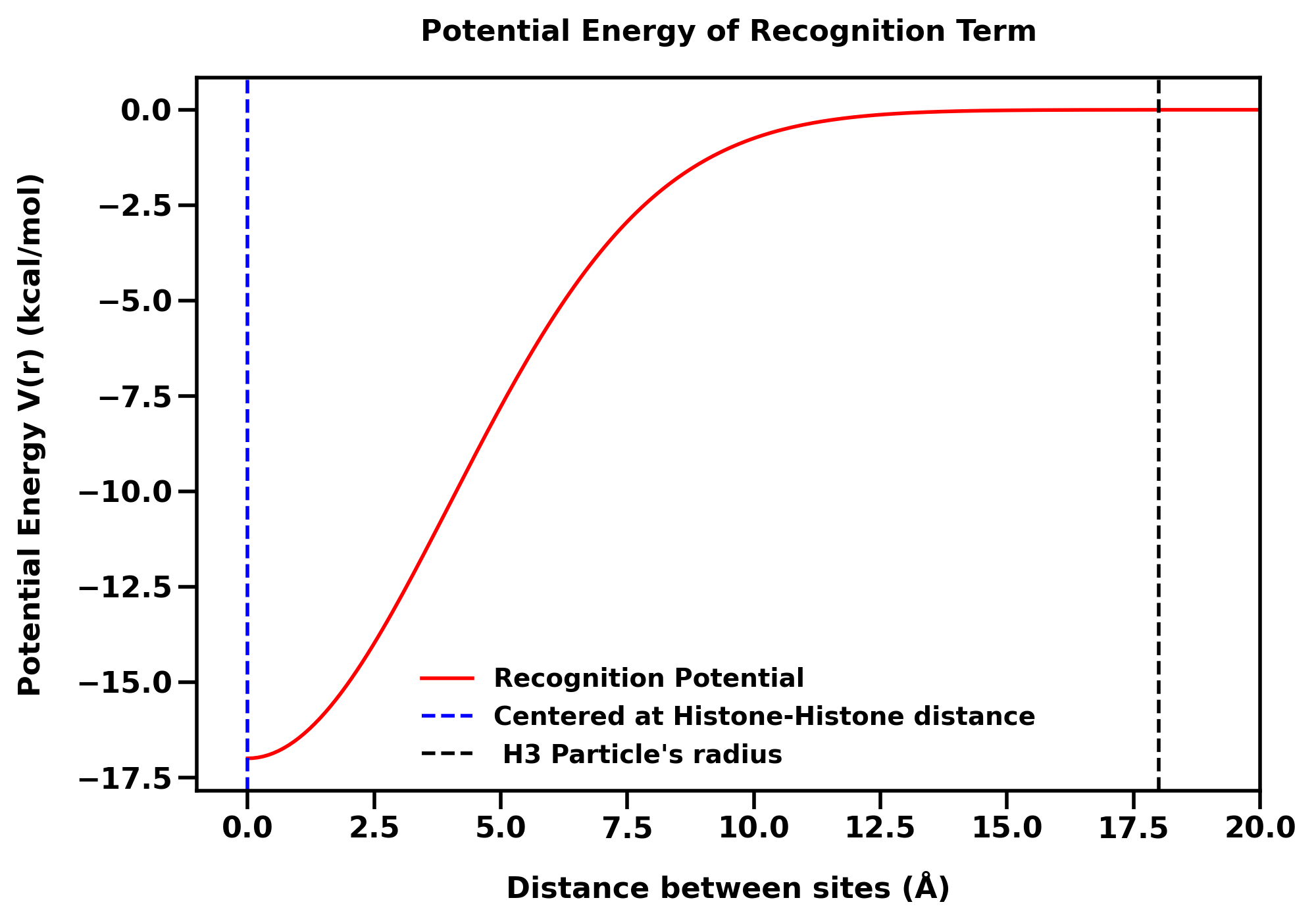}
\caption{Potential energy surface of the Recognition term.}
\label{fig:recog}
\end{figure}

\subsection*{Implementation}
We implemented our \textit{recognition}, attractive, and repulsive potentials in LAMMPS (Large-scale Atomic/Molecular Massively Parallel Simulator) \cite{thompson2022lammps}, a widely used open-source molecular dynamics package known for its parallel scalability and modular, object-oriented design. Specifically, we developed three custom pair classes within LAMMPS:
\begin{itemize}
\item \textbf{PairRec} encodes the off-center "Recognition" potential responsible for site-specific interactions between histone particles or histone-DNA binding sites (eq. \ref{eq:recog}).
\item \textbf{PairAttractive} handles the Gaussian-based attraction term (eq. \ref{eq:attract}).
\item \textbf{PairRepulsive} implements the exponential repulsion (eq. \ref{eq:repuls}).
\end{itemize}

Each class inherits from the core LAMMPS Pair base class, providing standard methods ($compute$, $settings$, $init\_style$, etc.) to calculate forces and potential energy for all interacting particle pairs at each timestep. This object-oriented structure enables a smooth integration with the rest of LAMMPS, so users can activate our potentials in an input script by specifying the $pair\_style$ keywords we provide.

We use LAMMPS’s fix brownian command that numerically implements the Euler--Maruyama scheme (eq. \ref{eq:EM}), discretizing the overdamped Langevin equation (eq. \ref{eq:brownian}) by combining deterministic drift and stochastic noise in a forward-time integration step. As a result, our code can run in parallel on multi-core CPUs without further modifications to the integrator, since LAMMPS handles domain decomposition \cite{plimpton1995fast} and data transfers internally. 

In addition to the CPU implementation, we developed a GPU-accelerated version of the model using the KOKKOS package \cite{edwards2014kokkos,trott2021kokkos} in LAMMPS. This involved re-writing all pair potentials in the $KOKKOS$ style and implementing a GPU-compatible $fix~brownian$ integrator that matches the standard CPU-based overdamped Langevin scheme. Our KOKKOS version enables deployment on GPU clusters and multi-GPU workstations without modifying user input files, offering a great speedup over the CPU version. This dual-mode support makes the force field accessible to both CPU and GPU users and ensures efficient scaling to large chromatin systems across high-performance computing resources.

By packaging our code into LAMMPS Pair classes, the entire force field is freely available to the community under LAMMPS’s open-source license. Researchers can directly compile these classes into their existing LAMMPS distributions, enabling large-scale simulations of not only histone-histone and histone-DNA, but also similarly site-specific protein-protein interactions. This parallel scalability and object-oriented architecture ensure that even million-base-pair or multi-nucleosome systems remain computationally tractable, facilitating a broad range of multiscale chromatin modeling studies

}

\showmatmethods{} 




\bibliography{references}

\begin{thebibliography}{10}

\bibitem{gardner2011operating}
KE Gardner, CD Allis, BD Strahl, Operating on chromatin, a colorful language
  where context matters.
\newblock {\em\protect\JournalTitle{Journal of molecular biology}}
  \textbf{409}, 36--46 (2011).

\bibitem{keenan2021discovering}
EK Keenan, DK Zachman, MD Hirschey, Discovering the landscape of protein
  modifications.
\newblock {\em\protect\JournalTitle{Molecular cell}} \textbf{81}, 1868--1878
  (2021).

\bibitem{black2012histone}
JC Black, C Van~Rechem, JR Whetstine, Histone lysine methylation dynamics:
  establishment, regulation, and biological impact.
\newblock {\em\protect\JournalTitle{Molecular cell}} \textbf{48}, 491--507
  (2012).

\bibitem{loh2012actors}
KM Loh, B Lim, Actors in the cell reprogramming drama.
\newblock {\em\protect\JournalTitle{Nature}} \textbf{488}, 599--600 (2012).

\bibitem{smith2013dna}
ZD Smith, A Meissner, Dna methylation: roles in mammalian development.
\newblock {\em\protect\JournalTitle{Nature Reviews Genetics}} \textbf{14},
  204--220 (2013).

\bibitem{bulger2011functional}
M Bulger, M Groudine, Functional and mechanistic diversity of distal
  transcription enhancers.
\newblock {\em\protect\JournalTitle{Cell}} \textbf{144}, 327--339 (2011).

\bibitem{hnisz2013super}
D Hnisz, et~al., Super-enhancers in the control of cell identity and disease.
\newblock {\em\protect\JournalTitle{Cell}} \textbf{155}, 934--947 (2013).

\bibitem{calo2013modification}
E Calo, J Wysocka, Modification of enhancer chromatin: what, how, and why?
\newblock {\em\protect\JournalTitle{Molecular cell}} \textbf{49}, 825--837
  (2013).

\bibitem{schaefer2023roles}
IM Schaefer, X Qian, The roles of the swi/snf complex in cancer.
\newblock {\em\protect\JournalTitle{Cancer cytopathology}} \textbf{131}, 410
  (2023).

\bibitem{zhao2021language}
S Zhao, CD Allis, GG Wang, The language of chromatin modification in human
  cancers.
\newblock {\em\protect\JournalTitle{Nature Reviews Cancer}} \textbf{21},
  413--430 (2021).

\bibitem{feng20213d}
Y Feng, X Liu, S Pauklin, 3d chromatin architecture and epigenetic regulation
  in cancer stem cells.
\newblock {\em\protect\JournalTitle{Protein \& Cell}} \textbf{12}, 440--454
  (2021).

\bibitem{apostolou2013chromatin}
E Apostolou, K Hochedlinger, Chromatin dynamics during cellular reprogramming.
\newblock {\em\protect\JournalTitle{Nature}} \textbf{502}, 462--471 (2013).

\bibitem{stevens20173d}
TJ Stevens, et~al., 3d structures of individual mammalian genomes studied by
  single-cell hi-c.
\newblock {\em\protect\JournalTitle{Nature}} \textbf{544}, 59--64 (2017).

\bibitem{tang2015ctcf}
Z Tang, et~al., Ctcf-mediated human 3d genome architecture reveals chromatin
  topology for transcription.
\newblock {\em\protect\JournalTitle{Cell}} \textbf{163}, 1611--1627 (2015).

\bibitem{markaki2012potential}
Y Markaki, et~al., The potential of 3d-fish and super-resolution structured
  illumination microscopy for studies of 3d nuclear architecture: 3d structured
  illumination microscopy of defined chromosomal structures visualized by 3d
  (immuno)-fish opens new perspectives for studies of nuclear architecture.
\newblock {\em\protect\JournalTitle{Bioessays}} \textbf{34}, 412--426 (2012).

\bibitem{ou2017chromemt}
HD Ou, et~al., Chromemt: Visualizing 3d chromatin structure and compaction in
  interphase and mitotic cells.
\newblock {\em\protect\JournalTitle{Science}} \textbf{357}, eaag0025 (2017).

\bibitem{lieberman2009comprehensive}
E Lieberman-Aiden, et~al., Comprehensive mapping of long-range interactions
  reveals folding principles of the human genome.
\newblock {\em\protect\JournalTitle{science}} \textbf{326}, 289--293 (2009).

\bibitem{lequieu20191cpn}
J Lequieu, A C{\'o}rdoba, J Moller, JJ De~Pablo, 1cpn: A coarse-grained
  multi-scale model of chromatin.
\newblock {\em\protect\JournalTitle{The Journal of chemical physics}}
  \textbf{150} (2019).

\bibitem{macpherson2018bottom}
Q MacPherson, B Beltran, AJ Spakowitz, Bottom--up modeling of chromatin
  segregation due to epigenetic modifications.
\newblock {\em\protect\JournalTitle{Proceedings of the National Academy of
  Sciences}} \textbf{115}, 12739--12744 (2018).

\bibitem{falk2019heterochromatin}
M Falk, et~al., Heterochromatin drives compartmentalization of inverted and
  conventional nuclei.
\newblock {\em\protect\JournalTitle{Nature}} \textbf{570}, 395--399 (2019).

\bibitem{bascom2018mesoscale}
GD Bascom, T Schlick, Mesoscale modeling of chromatin fibers in {\em Nuclear
  architecture and dynamics}.
\newblock (Elsevier), pp. 123--147 (2018).

\bibitem{brackley2016simulated}
CA Brackley, J Johnson, S Kelly, PR Cook, D Marenduzzo, Simulated binding of
  transcription factors to active and inactive regions folds human chromosomes
  into loops, rosettes and topological domains.
\newblock {\em\protect\JournalTitle{Nucleic acids research}} \textbf{44},
  3503--3512 (2016).

\bibitem{brandani2021kinetic}
GB Brandani, C Tan, S Takada, The kinetic landscape of nucleosome assembly: A
  coarse-grained molecular dynamics study.
\newblock {\em\protect\JournalTitle{PLoS computational biology}} \textbf{17},
  e1009253 (2021).

\bibitem{zhang2016exploring}
B Zhang, W Zheng, GA Papoian, PG Wolynes, Exploring the free energy landscape
  of nucleosomes.
\newblock {\em\protect\JournalTitle{Journal of the American Chemical Society}}
  \textbf{138}, 8126--8133 (2016).

\bibitem{laghmach2020mesoscale}
R Laghmach, M Di~Pierro, DA Potoyan, Mesoscale liquid model of chromatin
  recapitulates nuclear order of eukaryotes.
\newblock {\em\protect\JournalTitle{Biophysical Journal}} \textbf{118},
  2130--2140 (2020).

\bibitem{laghmach2021interplay}
R Laghmach, M Di~Pierro, DA Potoyan, The interplay of chromatin phase
  separation and lamina interactions in nuclear organization.
\newblock {\em\protect\JournalTitle{Biophysical Journal}} \textbf{120},
  5005--5017 (2021).

\bibitem{michieletto2018shaping}
D Michieletto, et~al., Shaping epigenetic memory via genomic bookmarking.
\newblock {\em\protect\JournalTitle{Nucleic acids research}} \textbf{46},
  83--93 (2018).

\bibitem{shi2018interphase}
G Shi, L Liu, C Hyeon, D Thirumalai, Interphase human chromosome exhibits out
  of equilibrium glassy dynamics.
\newblock {\em\protect\JournalTitle{Nature communications}} \textbf{9}, 3161
  (2018).

\bibitem{verdaasdonk2011centromeres}
JS Verdaasdonk, K Bloom, Centromeres: unique chromatin structures that drive
  chromosome segregation.
\newblock {\em\protect\JournalTitle{Nature reviews Molecular cell biology}}
  \textbf{12}, 320--332 (2011).

\bibitem{vasquez2016entropy}
PA Vasquez, et~al., Entropy gives rise to topologically associating domains.
\newblock {\em\protect\JournalTitle{Nucleic Acids Research}} \textbf{44},
  5540--5549 (2016).

\bibitem{gursoy2016three}
G G{\"u}rsoy, J Liang, Three-dimensional chromosome structures from energy
  landscape.
\newblock {\em\protect\JournalTitle{Proceedings of the National Academy of
  Sciences}} \textbf{113}, 11991--11993 (2016).

\bibitem{buckle2018polymer}
A Buckle, CA Brackley, S Boyle, D Marenduzzo, N Gilbert, Polymer simulations of
  heteromorphic chromatin predict the 3d folding of complex genomic loci.
\newblock {\em\protect\JournalTitle{Molecular cell}} \textbf{72}, 786--797
  (2018).

\bibitem{michieletto2016polymer}
D Michieletto, E Orlandini, D Marenduzzo, Polymer model with epigenetic
  recoloring reveals a pathway for the de novo establishment and 3d
  organization of chromatin domains.
\newblock {\em\protect\JournalTitle{Physical Review X}} \textbf{6}, 041047
  (2016).

\bibitem{kang2015confinement}
H Kang, YG Yoon, D Thirumalai, C Hyeon, Confinement-induced glassy dynamics in
  a model for chromosome organization.
\newblock {\em\protect\JournalTitle{Physical review letters}} \textbf{115},
  198102 (2015).

\bibitem{arya2006role}
G Arya, T Schlick, Role of histone tails in chromatin folding revealed by a
  mesoscopic oligonucleosome model.
\newblock {\em\protect\JournalTitle{Proceedings of the National Academy of
  Sciences}} \textbf{103}, 16236--16241 (2006).

\bibitem{ozer2015chromatin}
G Ozer, A Luque, T Schlick, The chromatin fiber: multiscale problems and
  approaches.
\newblock {\em\protect\JournalTitle{Current opinion in structural biology}}
  \textbf{31}, 124--139 (2015).

\bibitem{mendenhall2013locus}
EM Mendenhall, et~al., Locus-specific editing of histone modifications at
  endogenous enhancers.
\newblock {\em\protect\JournalTitle{Nature biotechnology}} \textbf{31},
  1133--1136 (2013).

\bibitem{benveniste2014transcription}
D Benveniste, HJ Sonntag, G Sanguinetti, D Sproul, Transcription factor binding
  predicts histone modifications in human cell lines.
\newblock {\em\protect\JournalTitle{Proceedings of the National Academy of
  Sciences}} \textbf{111}, 13367--13372 (2014).

\bibitem{stanton2018chemically}
BZ Stanton, EJ Chory, GR Crabtree, Chemically induced proximity in biology and
  medicine.
\newblock {\em\protect\JournalTitle{Science}} \textbf{359}, eaao5902 (2018).

\bibitem{zhou2024induced}
C Zhou, S Wagner, FS Liang, Induced proximity labeling and editing for
  epigenetic research.
\newblock {\em\protect\JournalTitle{Cell chemical biology}} \textbf{31},
  1118--1131 (2024).

\bibitem{hathaway2012dynamics}
NA Hathaway, et~al., Dynamics and memory of heterochromatin in living cells.
\newblock {\em\protect\JournalTitle{Cell}} \textbf{149}, 1447--1460 (2012).

\bibitem{chiarella2020dose}
AM Chiarella, et~al., Dose-dependent activation of gene expression is achieved
  using crispr and small molecules that recruit endogenous chromatin machinery.
\newblock {\em\protect\JournalTitle{Nature biotechnology}} \textbf{38}, 50--55
  (2020).

\bibitem{armeev2021histone}
GA Armeev, AS Kniazeva, GA Komarova, MP Kirpichnikov, AK Shaytan, Histone
  dynamics mediate dna unwrapping and sliding in nucleosomes.
\newblock {\em\protect\JournalTitle{Nature communications}} \textbf{12}, 2387
  (2021).

\bibitem{kabsch1976solution}
W Kabsch, A solution for the best rotation to relate two sets of vectors.
\newblock {\em\protect\JournalTitle{Foundations of Crystallography}}
  \textbf{32}, 922--923 (1976).

\bibitem{bucciarelli2016dramatic}
S Bucciarelli, et~al., Dramatic influence of patchy attractions on short-time
  protein diffusion under crowded conditions.
\newblock {\em\protect\JournalTitle{Science advances}} \textbf{2}, e1601432
  (2016).

\bibitem{zhu2020self}
Y Zhu, A Bansal, S Xi, J Lu, WG Chapman, Self-assembly and phase behavior of
  mixed patchy colloids with any bonding site geometry: theory and simulation.
\newblock {\em\protect\JournalTitle{Soft Matter}} \textbf{16}, 3806--3820
  (2020).

\bibitem{kraft2012surface}
DJ Kraft, et~al., Surface roughness directed self-assembly of patchy particles
  into colloidal micelles.
\newblock {\em\protect\JournalTitle{Proceedings of the National Academy of
  Sciences}} \textbf{109}, 10787--10792 (2012).

\bibitem{tramantano2016constitutive}
M Tramantano, et~al., Constitutive turnover of histone h2a. z at yeast
  promoters requires the preinitiation complex.
\newblock {\em\protect\JournalTitle{Elife}} \textbf{5}, e14243 (2016).

\bibitem{niina2017sequence}
T Niina, GB Brandani, C Tan, S Takada, Sequence-dependent nucleosome sliding in
  rotation-coupled and uncoupled modes revealed by molecular simulations.
\newblock {\em\protect\JournalTitle{PLoS computational biology}} \textbf{13},
  e1005880 (2017).

\bibitem{bowman2014post}
GD Bowman, MG Poirier, Post-translational modifications of histones that
  influence nucleosome dynamics.
\newblock {\em\protect\JournalTitle{Chemical reviews}} \textbf{115}, 2274--2295
  (2014).

\bibitem{hazan2015nucleosome}
NP Hazan, et~al., Nucleosome core particle disassembly and assembly kinetics
  studied using single-molecule fluorescence.
\newblock {\em\protect\JournalTitle{Biophysical journal}} \textbf{109},
  1676--1685 (2015).

\bibitem{nawara2022imaging}
TJ Nawara, et~al., Imaging vesicle formation dynamics supports the flexible
  model of clathrin-mediated endocytosis.
\newblock {\em\protect\JournalTitle{Nature communications}} \textbf{13}, 1732
  (2022).

\bibitem{zhang2016rapid}
Z Zhang, et~al., Rapid dynamics of general transcription factor tfiib binding
  during preinitiation complex assembly revealed by single-molecule analysis.
\newblock {\em\protect\JournalTitle{Genes \& development}} \textbf{30},
  2106--2118 (2016).

\bibitem{banani2017biomolecular}
SF Banani, HO Lee, AA Hyman, MK Rosen, Biomolecular condensates: organizers of
  cellular biochemistry.
\newblock {\em\protect\JournalTitle{Nature reviews Molecular cell biology}}
  \textbf{18}, 285--298 (2017).

\bibitem{kong2009using}
S Kong, AF Day, RD O'Kennedy, PA Shamlou, NJ Titchener-Hooker, Using
  viscosity-time plots of escherichia coli cells undergoing chemical lysis to
  measure the impact of physiological changes occurring during batch cell
  growth.
\newblock {\em\protect\JournalTitle{Journal of Chemical Technology \&
  Biotechnology: International Research in Process, Environmental \& Clean
  Technology}} \textbf{84}, 696--701 (2009).

\bibitem{shaytan2016coupling}
AK Shaytan, et~al., Coupling between histone conformations and dna geometry in
  nucleosomes on a microsecond timescale: atomistic insights into nucleosome
  functions.
\newblock {\em\protect\JournalTitle{Journal of molecular biology}}
  \textbf{428}, 221--237 (2016).

\bibitem{kroon1997estimation}
LM Kroon-Batenburg, PH Kruiskamp, JF Vliegenthart, J Kroon, Estimation of the
  persistence length of polymers by md simulations on small fragments in
  solution. application to cellulose.
\newblock {\em\protect\JournalTitle{The Journal of Physical Chemistry B}}
  \textbf{101}, 8454--8459 (1997).

\bibitem{mardian1978yeast}
JK Mardian, I Isenberg, Yeast inner histones and the evolutionary conservation
  of histone-histone interactions.
\newblock {\em\protect\JournalTitle{Biochemistry}} \textbf{17}, 3825--3833
  (1978).

\bibitem{thompson2022lammps}
AP Thompson, et~al., Lammps-a flexible simulation tool for particle-based
  materials modeling at the atomic, meso, and continuum scales.
\newblock {\em\protect\JournalTitle{Computer physics communications}}
  \textbf{271}, 108171 (2022).

\bibitem{plimpton1995fast}
S Plimpton, Fast parallel algorithms for short-range molecular dynamics.
\newblock {\em\protect\JournalTitle{Journal of computational physics}}
  \textbf{117}, 1--19 (1995).

\bibitem{edwards2014kokkos}
HC Edwards, CR Trott, D Sunderland, Kokkos: Enabling manycore performance
  portability through polymorphic memory access patterns.
\newblock {\em\protect\JournalTitle{Journal of parallel and distributed
  computing}} \textbf{74}, 3202--3216 (2014).

\bibitem{trott2021kokkos}
CR Trott, et~al., Kokkos 3: Programming model extensions for the exascale era.
\newblock {\em\protect\JournalTitle{IEEE Transactions on Parallel and
  Distributed Systems}} \textbf{33}, 805--817 (2021).

\end{thebibliography}

\end{document}